**A Genetic Algorithm Trained Machine-Learned Interatomic Potential for the Silicon-Carbon System**

Michael MacIsaac[a], Salil Bavdekar[b], Douglas Spearot[a], Ghatu Subhash[a]

[a] Department of Mechanical and Aerospace Engineering, University of Florida, Gainesville, FL 32611, USA

[b] Department of Material Science and Engineering, University of Florida, Gainesville, FL 32611, USA

## Abstract

A linear regression-based machine learned interatomic potential (MLIP) was developed for the silicon-carbon system. The MLIP was predominantly trained on structures discovered through a genetic algorithm, encompassing the entire silicon-carbon composition space, and uses as its foundation the Ultra-Fast Force Fields (UF3) formulation. To improve MLIP performance, the learning algorithm was modified to include higher spline interpolation resolution in regions with large potential energy surface curvature. The developed MLIP demonstrates exceptional predictive performance, accurately estimating energies and forces for structures across the silicon-carbon composition and configuration space. The MLIP predicts mechanical properties of silicon carbide (SiC) with high precision and captures fundamental volume-pressure and volume-temperature relationships. Uniquely, this silicon-carbon MLIP is adept at modeling complex high-temperature phenomena, including the peritectic decomposition of SiC and carbon dimer formation during SiC surface reconstruction, which cannot be captured with prior classical interatomic potentials for this material.

## 1. Introduction

Traditional approaches to understand and predict material behavior involve experimentation and development of constitutive models. However, these approaches may not yield insight into atomic scale deformation mechanisms like bond breaking and phase transformations. Consequently, computational tools have been developed to gain greater understanding at the atomic level. These computational approaches fall primarily into two categories: (i) *ab initio* methods based on density functional theory (DFT) [1] and (ii) classical atomistic simulations using an interatomic potential [2,3]. *Ab initio* methods employ a quantum mechanical approach to model the electronic structure and compute forces in many-body systems. However, these approaches come with great computational costs – scaling approximately as $O(N^3)$ with N being the number of atoms – limiting their domain to a few hundred atoms and a few picoseconds simulation time [4]. Conversely, classical atomistic simulations are more computationally efficient [5] because atoms are treated as point masses, and the electronic degrees of freedom are not explicitly considered. This efficiency affords time and length scales orders of magnitude larger than DFT for the same computational resources, enabling the modeling of phenomena like shock [6] and phase transitions [7]. However, access to these time and length scales may come with reduced accuracy depending on the quality of the interatomic potential used to compute atomic forces and energies.

In recent years, machine learned interatomic potentials (MLIPs) have grown in popularity as they show increasing promise to narrow the gap in accuracy between *ab initio* calculations and classical atomistic simulations performed with traditional interatomic potentials [8–11] . Improvements can be attributed to the use of machine learning (ML) methodologies (e.g., neural networks (NN)) to approximate the potential energy surface (PES) of a material and learn fundamental structure-energy relationships that govern material performance. This understanding constitutes the foundation of MLIP property prediction.

Generally, MLIP development requires three main components: (i) training data, (ii) material structure descriptor, and (iii) learning algorithm. Typically, the training data is generated through DFT calculations, involving snapshots of static structures and/or snapshots extracted from *ab initio* molecular dynamics (AIMD) trajectories.  Data generation approaches must adequately sample the PES and domain of interest to ensure desired MLIP predictive performance for a wide range of atomic environments. Moreover, the generation of training data can be a product of manual selection [12], based on domain expertise, automated procedures such as active-learning [13,14], and/or genetic algorithms (GAs) [8,15,16] (as employed in this study). Next, the material structure descriptor encodes the training data into a machine learnable input [9–11]. Descriptors describe atomic environments with varying degrees of complexity and in turn impact the computational cost of a MLIP. Furthermore, descriptors feature adjustable parameters, which should be optimized to strike a balance between accuracy and computational cost.  The final component of a MLIP is the learning algorithm. Popular learning algorithms include neural networks (NN) [17–19], kernel methods [20–23], and linear regression [10,14,24,25]. Neural networks are popular due to their flexible functional form and ability to serve as universal approximators [26]. However, their flexibility necessitates notably more training data to safeguard against overfitting in comparison to other ML algorithms [5]. Additionally, the large number of fitting parameters in NNs can lead to higher computational costs as compared to other approaches. Kernel methods, such as Gaussian regression and kernel ridge regression, rely on similarity measures between a configuration and the training data, correlating the MLIP's computational cost with dataset size. This correlation, as seen with Gaussian

Approximation Potentials [20], can result in significant computational expense. Linear regression [10,14,24] approaches assume a linear relationship between energy and descriptor components, offering simplicity as well as greater computational efficiency than NN and kernel-based approaches. However, linear regression potentials may experience increased error as compared to a NN. This underperformance can be attributed to a more rigid functional form.

To date, MLIPs have been developed predominately for single element metallic materials, such as tungsten [27,28], iron [29,30], aluminum [31,32], etc. While a few models for ceramics have been developed [33–35], they are not nearly as prevalent. Ceramics possess attributes such as high hardness, chemical inertness, low density, high temperature resistance, oxidation and corrosion resistance, etc., that are well suited for use in harsh environments. Consequently, precise modeling is critical for studying their behavior under extreme conditions. However, modeling ceramics poses challenges due to their multi-element structure and intricate crystal geometry, characterized by highly directional covalent bonds.

A ceramic of increasing importance is silicon carbide (SiC) which exhibits extreme polymorphism (over 250 polytypes) and demonstrates high-performance under extreme conditions such as ballistic impact [36–39], high temperature [40], and under irradiation [41,42]. Moreover, high temperature and oxidation resistance has promoted its use in hypersonic [43–46], and nuclear power [47–49] applications. Numerous computational studies [50–57] via classical atomistic simulations or *ab initio* methods have been performed on SiC to better understand its deformation behavior and phase transitions. A popular empirical potential for studying SiC behavior is the Vashishta potential [58], which has been used to investigate SiC shock response with system sizes exceeding six million atoms [53]. This potential predicts fitted properties such as cohesive energy, bulk modulus, and the $C_{11}$ elastic constant with a high degree of accuracy, but fails to accurately reproduce other properties, such as shear moduli and the $C_{33}$ elastic constant in the SiC-3C, SiC-2H, and SiC-6H polytypes (as will be discussed later in this work). Moreover, the Vashishta potential was not formulated to capture the physics of same atom (Si-Si or C-C) interactions [58]. This limitation is detrimental when evaluating non-equimolar stoichiometries, which may arise during simulations involving defects [59], as well as during thermal decomposition, where a peritectic reaction involving silicon sublimation is reported [60–63].

The objective of this work is to develop a new interatomic potential for the Si-C system, using a machine learning method, that narrows the gap between *ab initio* accuracy and classical molecular dynamics efficiency. More specifically, the aim is to capture the physics of the Si-C system across the composition and temperature space, providing a simulation tool that will elucidate greater knowledge of SiC behavior in extreme environments. The training data was primarily generated via a GA search over the entire Si-C compositional space, encompassing equimolar and non-equimolar compositions. The GA was used to identify energetically favorable structures, maximize dataset diversity, and extend model applicability beyond known configurations and equimolar compositions. The model was trained using the Ultra-Fast Force Fields (UF$^3$) development package [24], which employs a linear regression-based approach for PES modeling. The choice of linear regression over learning algorithms such as NNs was influenced by the large quantity of training data that would be necessary to adequately sample the Si-C system if using NNs. Furthermore, the versatility of linear regression was seen as advantageous for modeling SiC polymorphism. Modifications to the learning algorithm were made which led to significant gains in model predictive performance. The developed MLIP has force and energy errors comparable to single element metallic materials, predicts mechanical properties with a high degree of accuracy, and is adept at modeling complex phenomena not explicitly included in the training data.

## 2. Methods

In this work, the process of developing the MLIP consisted of four phases: training data generation, data featurization, model training, and model validation. These tasks were tailored for the Si-C system and the primary goal of this section is to clearly document the MLIP development approach employed.

### 2.1 Training data

The training data was primarily generated via the Genetic Algorithm for Structure and Phase Prediction (GASP) software package [15]. A GA is a heuristic optimization method that adopts an evolutionary approach by employing variation operators like selection, crossover, and mutation, to evolve a population of candidate solutions towards a near-optimal solution for a given problem [64]. As applied to materials, a GA is seeded with initial (parent) structures and explores a material composition space via iterative hybridization to identify energetically favorable structures.

GAs are favorable for dataset generation as they contribute to greater dataset diversity [65,66], providing a more complete representation of the PES. Importantly, GAs are free from biases, such as physical intuition, which may be present in approaches like AIMD sampling. The GASP generated data includes identified structures and their relaxation trajectories. Unrelaxed structures were included to enrich the dataset diversity and further define the PES via the inclusion of higher energy structures [65]. Due to strong correlations between structures from the same relaxation trajectory, relaxation trajectories were exclusively designated to either training or testing data.

The GASP runs explored the entire Si-C composition space, encompassing equimolar and non-equimolar compositions and inclusive of the Si and C end points. Figure 1 demonstrates the diversity of structures identified through the GASP search, highlighting the substantial prevalence of non-equimolar structures. The structure searches were seeded with multiple known polytypes (e.g., SiC-3C, SiC-2H, SiC-4H, SiC-6H, SiC-15R, etc.), as well as Si and C lattices. In addition to providing parent structures, search settings, including structure constraints (e.g., minimum and maximum number of atoms), variation operator constraints (e.g., probability of structure mating versus mutation versus permutation), and *ab initio* calculation settings (e.g., convergence criteria), must be specified. Information related to the GASP search parameters can be found on GitHub (https://github.com/SubhashUFlorida/SiC-MLIP).

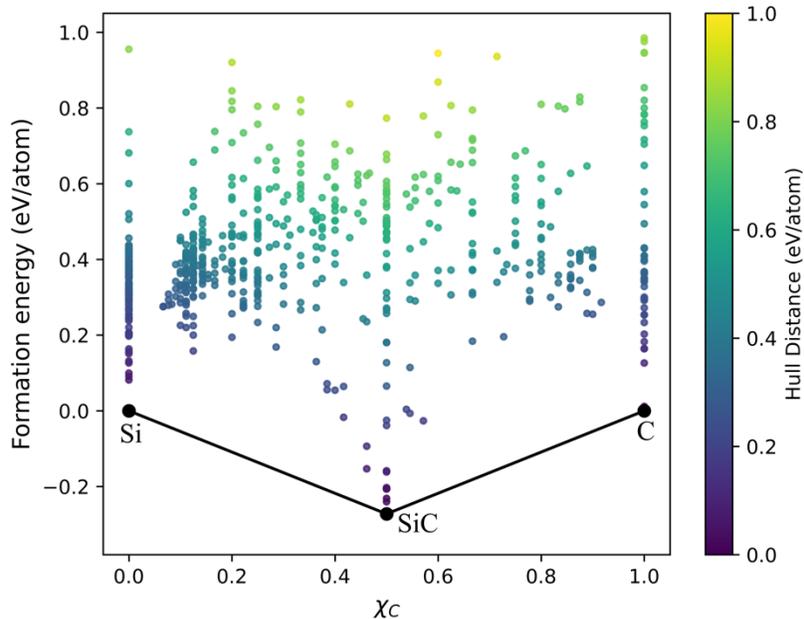

Figure 1: GASP identified structures within the Si-C composition space, where a structure's energy (stability) above hull is indicated by the intensity of its diamond marker. Only final (energy minimized) structures are shown.

While no stable non-equimolar Si-C polytype is known to exist, a design space phase search (no stoichiometric limitations) was performed. The aim was to maximize local atomic environment diversity and condition the MLIP for situations where non-equimolar compositions may arise, such as melting and other phase transformations.

To improve bulk moduli prediction, *ab* initio molecular dynamics (AIMD) [67] sampling was employed to supplement the dataset with three pertinent polytypes: SiC-3C (zinc-blende and rock salt structures), SIC-2H, and SiC-6H. These polytypes were emphasized due to their stability and/or prevalence in extreme applications. Ultimately, the training data included 41225 structures, where 41145 structures were generated via GASP and the remaining 80 were collected via AIMD sampling.

All DFT calculations were performed using the Vienna Ab initio Simulation Package [68] using the Perdew-Burke-Ernzerhof (PBE) functional. KPOINTS were automatically generated with a gamma centered mesh and length of 30 (used for subdivision determination). The structural relaxations were carried out with tolerances of $10^{-6}$ eV for electronic convergence and $10^{-5}$ eV for ionic convergence, and an energy cutoff of 520 eV. For further information related to VASP calculation details see the data folder on the GitHub.

## 2.2 Data featurization

Following data generation, the training data underwent a featurization process as dictated by the $UF^3$ descriptor [24]. The goal of the $UF^3$ descriptor is to efficiently capture the PES through a low-order many-body expansion, while accurately describing energies, forces, and phonon frequencies. This many-body expansion is limited to two- and three-body terms, with dependencies on one and three pairwise distances, respectively. These two- and three-body terms are represented via a collection of cubic B-spline basis functions. The local atomic environment encoded by the $UF^3$ descriptor is predominantly dictated by the number of cubic B-splines and the specified minimum and maximum cutoff radii over which the local atomic environment is defined. Cubic B-splines are advantageous due to their rapid evaluation, continuity, and smoothness, particularly for the first order derivative and continuous nature for the second order derivative, allowing them to capture phonon frequencies.

In spline interpolation, the splines are connected at knot positions. In the case of $UF^3$, the splines provide compact support, where they are non-zero over four adjacent knot spacings. This compact support promotes the computational efficiency of $UF^3$, requiring only the evaluation of 4 or 64 basis functions to calculate any two or three body term, respectively. Furthermore, the endpoints are padded with additional basis functions. This adjustment ensures that at short distances near the minimum cutoff radii, interactions are strongly repulsive to prevent atoms from becoming unphysically close. Conversely, at greater distances near the maximum cutoff radii, the potential smoothly diminishes to zero.

In this work, the minimum and maximum cutoff radii of the descriptor were determined via a convergence study that evaluated energy and force errors as a function of different cutoff values. The study was conducted with minimum cutoff radii for two- and three-body terms of 0.001 Å and (0.5, 0.5, 0.5) Å, respectively. Three body cutoffs are specified as a triplet of distance because a three-body term involves three pair interactions. The two-body minimum cutoff radii were set less than the three-body minimum to account for close interactions that may arise under high temperature conditions. Additionally, a requirement of this study was that the maximum cutoff radius for two-body interactions must be greater than or equal to those for three-body interactions. This condition was set due to the predominant influence of two-body contributions over three-body interactions and aims to reduce computational cost, where three-body terms are the dominant contributors to computational cost. The force error converged when the three-body maximum cutoff was (5.0, 5.0, 10.0) Å, hence the three-body minimum and maximum radial cutoffs were (0.5, 0.5, 0.5) Å and (5.0, 5.0, 10.0) Å, respectively. Thus, the two-body (pairwise interactions) were encoded with a minimum and maximum radial cutoff of 0.001 and 6.0 angstroms, respectively. Figures that illustrate these findings and support the final decision on cutoffs can be found in the supplemental information on the GitHub.

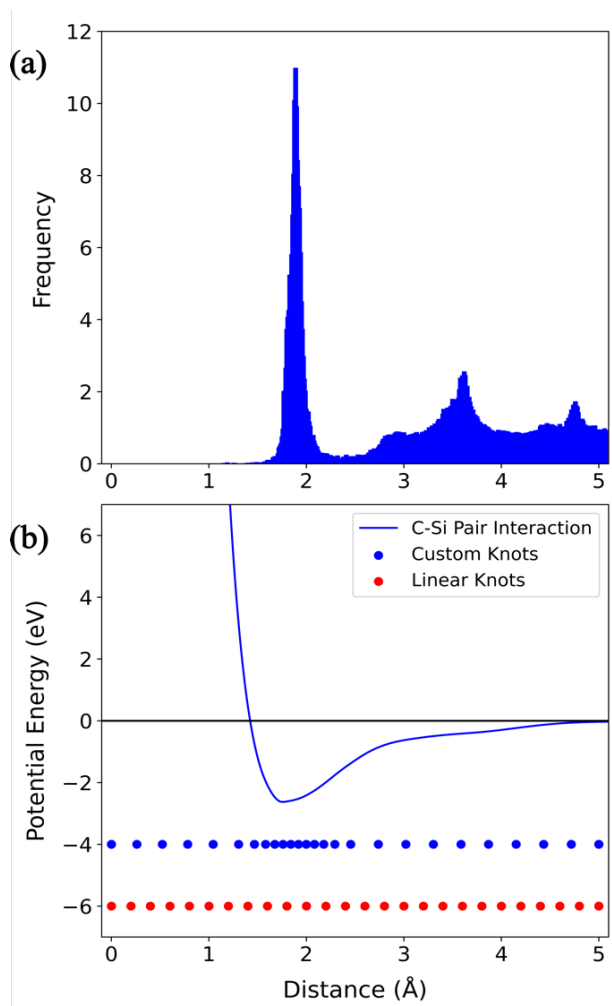

Figure 2: (a) Radial distribution function of C-Si pair interactions for the entire training dataset, and (b) C-Si pair interaction and knot locations using both the linear strategy (red dots) and the custom strategy (blue dots).

In UF$^3$, the default knot spacing for the cubic B-splines is linear. However, linear spacing may not be the optimal strategy for knot placement. Upon examining the radial distribution function of C-Si pair interaction distances from the entire training dataset (Fig. 2a), a distinct peak occurs at ~1.90 angstroms, corresponding to the energy well minima of the pair interaction curve (Fig 2b). Thus, a custom knot strategy was developed featuring a fixed-width, gaussian-like distribution of knots centered around the energy well minima, with adjacent linear regions. To highlight the differences in knot strategies, the knot positions as dictated by the custom and linear strategies are represented as blue and red dots, respectively, in Fig. 2b. It is important to note that an equal number of knots are plotted for both the linear and custom knot strategies. The rationale behind the custom strategy was to increase knot resolution in regions of maximum PES curvature and high data density. Increased knot resolution correlates to more model fitting parameters, providing greater flexibility in capturing energy well curvature. Furthermore, this custom strategy reduces the probability of overfitting as there are fewer knots in the adjacent linear-spaced regions.

To validate the custom knot strategy, grid searches were performed for both strategies. Both searches explored the same model hyperparameter space, and each model's performance was evaluated using error metrics and material property prediction. During model fitting, UF$^3$ employs Tikhonov regularization [69] with five regularizer hyperparameters encompassing one, two, and three body ridge regularizers, as well as two and three body curvature regularizers. Ridge regularizers aim to reduce overfitting by penalizing large spline coefficients, while curvature regularizers control the curvature and smoothness across neighboring spline coefficients. Additionally, UF$^3$ employs an energy weight term that dictates the degree to minimize energy error relative to force error.

The grid searches utilized a coarse grid of hyperparameter values. Regularization parameter values ranged from $10^{-2}$ to $10^{-12}$ in decrements of two orders of magnitude, while the energy/force weight parameter ranged from 0.05 to 1.0 in increments of 0.05. To quantitatively assess the efficacy of the two strategies, performance benchmarks were established, including normalized error as defined in Eq. (1) and material property predictions using the Large-scale Atomic/Molecular Massively Parallel Simulator (LAMMPS) [70]. The normalized error is the mean of the root mean square energy and force errors ($RMSE_E$ and $RMSE_F$), normalized by their respective standard deviations ($\sigma_E$ and $\sigma_F$),

$$\text{Normalized Error} = \frac{1}{2}\left(\frac{RMSE_E}{\sigma_E} + \frac{RMSE_F}{\sigma_F}\right) \ . \tag{1}$$

For a model to be considered acceptable it must pass the benchmarks shown in Table I.

**Table I: Knot strategy performance benchmarks**

| Benchmark | Requirement |
|---|---|
| Normalized error | < 0.20 |
| Cohesive energy error (SiC-3C, SiC-6H) | < 10% |
| Lattice constant error (SiC-3C) | < 5% |
| Bulk and shear moduli error (SiC-3C, SiC-6H) | < 20% |
| Bulk moduli – pressure relationship (SiC-3C) | Positive |

The listed material property benchmarks were selected as they are fast to evaluate and provide insight into the applicability and physical interpretability of trained models. These material properties were evaluated for the SiC-3C and SiC-6H polytypes due to their prevalence in the literature and distinct features. More specifically, SiC-3C has a cubic crystal structure with carbon coordinated atoms, whereas SiC-6H has a hexagonal crystal structure with either cubic or hexagonal coordinated carbon. Metric requirements were reflective of their importance. Lattice constant prediction was given the least margin of error, as it directly influences density and elastic properties. Furthermore, low cohesive energy error is crucial for accurately capturing the ground state. In contrast, bulk and shear moduli errors were the least stringent, as they are more extrapolative than the previous two.

Among the evaluated models, no models trained using the linear knot strategy met the full set of criteria, while 162 models trained using the custom knot strategy were deemed acceptable. The linear knot strategy may achieve success in individual metrics, but cannot simultaneously pass all set targets in Table 1, in contrast to the custom knot strategy which proves capable of meeting these requirements. Hence, the custom knot strategy, newly proposed in this work, can more effectively capture the underlying physics. Due to its marked performance improvements over the linear knot strategy, the custom knot strategy has been formalized and is now available in UF$^3$ (https://github.com/uf3/uf3).

### 2.3 Model fitting and validation

To determine the optimal hyperparameter combination, the custom knot strategy grid search was refined to explore more effective hyperparameter domains that emerged during the initial grid search. Models demonstrating high performance across preliminary metrics (Table I) were chosen for further validation using large deformations to predict the change in bulk modulus with pressure ($P$),

$$K(P) = K_0 + K_0'P \tag{2}$$

where $K_0$ is the bulk modulus at the ground state ($P = 0$) and $K_0'$ is the derivative of bulk modulus with respect to pressure. The ground state structure of SiC-3C was deformed up to ±10% volumetric strain, the resulting energy and pressures were calculated using DFT and the MLIP via LAMMPS simulations. The data was fit using the Murnaghan equation of state (EOS) [71],

$$P(V) = \frac{K_o}{K_o'}\left[\left(\frac{V}{V_o}\right)^{-K_o'} - 1\right] \tag{3}$$

Here $V_0$ is the initial volume of the structure (in the ground state), and $P$ is the pressure of the structure when compressed to a final volume $V$. Models were filtered to include those with $K_0'$ values close to that of the DFT calculations ($K_0' = 3.76$).

Acceptable models underwent further validation by comparing their volume-pressure and volume-temperature relationships with *ab initio* findings. Associated simulation data were used to fit a volume-pressure EOS and a volume-temperature EOS as

$$V(P) = V_0 + V_0'P + V_0''P^2 \tag{4}$$

$$V(T) = V_0 + V_0'T + V_0''T^2 \tag{5}$$

Models whose overall volume-pressure and volume-temperature relationships aligned with *ab initio*, were then evaluated by comparing their pair interaction energy curves to those produced by DFT. Pairwise interactions are another physically interpretable metric and are the dominant contributors to the total energy and properties of the system. Therefore, MLIPs used for meaningful simulations should resemble DFT predictions (e.g., relative depth of pair interaction energy wells).

Finally, the highest-performing models up to this point were chosen for melting temperature prediction simulations. Melting simulations were conducted as the last test for two main reasons: (i) they are the most computationally demanding, requiring orders of magnitude longer simulation times compared to previous tests, and (ii) they evaluate an MLIP's stability under extreme conditions. More specifically, we seek a MLIP that effectively models high energy atomic configurations, including situations not explicitly included in the training data, such as solid-liquid interfaces during melting.

The melting temperature simulations were carried out using the coexistence method [72], which to our knowledge has not been extensively used for ceramics. In brief, the coexistence method involves dividing a simulation cell in half, where the bottom half is heated to an equilibrium temperature, denoted as $T_E$, and maintained as a solid, while the top half is heated well above the melting temperature so that it transitions into a liquid, and then quenched to $T_E$. Subsequently, the two halves are brought together and allowed to reach thermal equilibrium. The interface between the solid and liquid halves serves as a nucleation site, facilitating the growth of either the solid or melted phase depending on the temperature. If the specified $T_E$ is below the melting temperature, the solid region will expand, whereas if $T_E$ is above the melting temperature, the solid region will contract. The coexistence method offers advantages over homogenous nucleation methods, which often lead to superheated structures and consequently overpredict the melting temperature.

The MLIP which displayed the greatest holistic performance across all metrics described thus far is chosen as the final MLIP and its performance is evaluated in the following sections. The data used to train the model, python scripts used during featurization and MLIP training, as well as the final model can be found on GitHub.

## 3. Results & Discussion

In this section, we highlight the performance of the developed Si-C MLIP and draw comparisons with existing literature. When making comparisons, we reference literature values when available. In other cases, comparisons are made to DFT calculations and/or LAMMPS simulations using the Vashishta potential [58]. As described above, the MLIP was primarily evaluated on its predictions for the mechanical and thermal properties of interest on the most prevalent SiC polytypes.

## 3.1 Learning energy and force errors

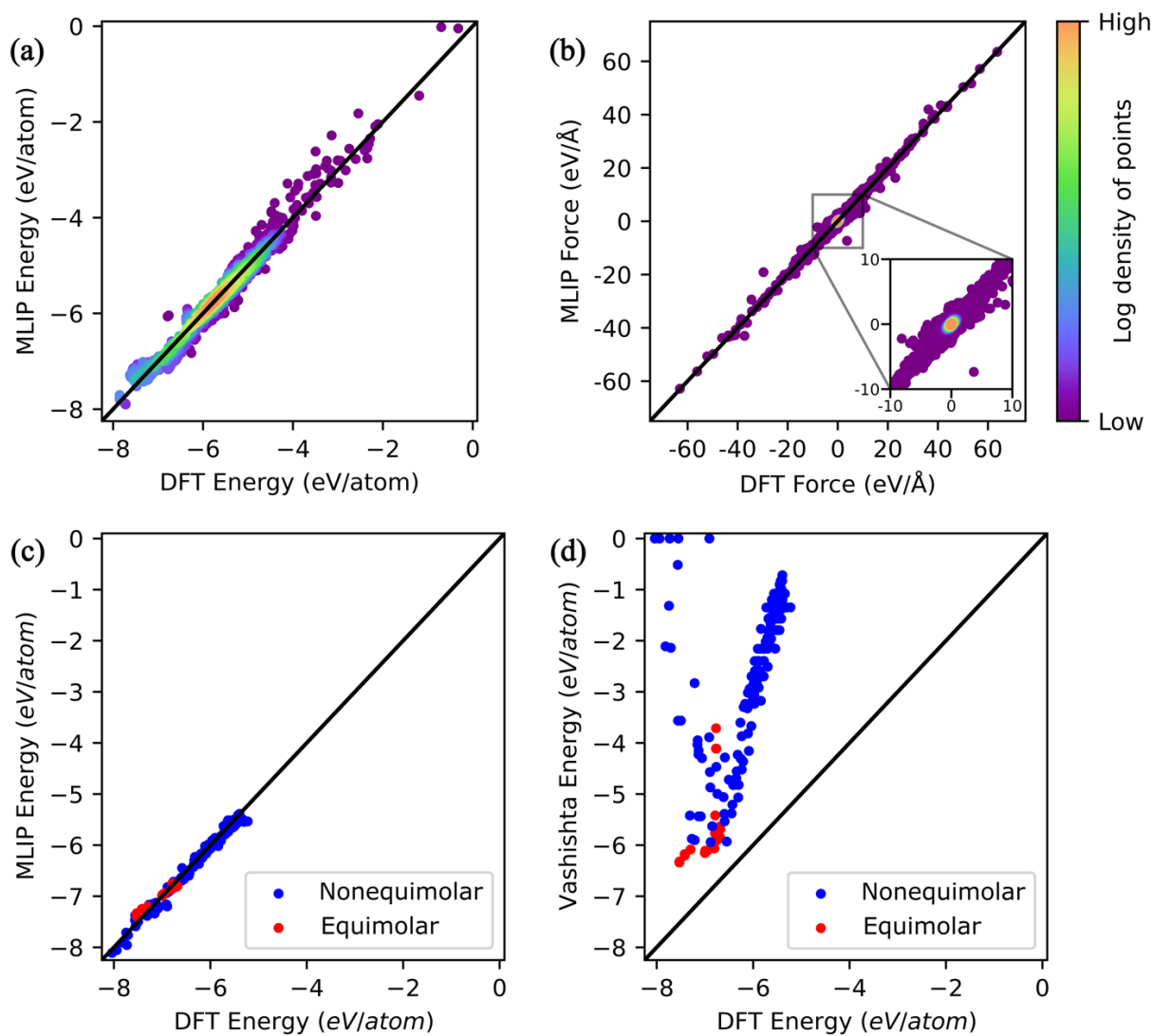

Figure 3: MLIP predicted versus reference (a) energies and (b) forces (inset is a close up between -10 and +10 meV/Å). Comparison of LAMMPS energy evaluations using the (c) MLIP and (d) Vashishta potential for ~200 structures with varied compositions against reference DFT values.

The training data diversity is shown in Figs. 3(a) and 3(b), where the MLIP predicted values for energies and forces are plotted against the DFT values. Predictions using the MLIP align well with the DFT reference data over energy values ranging from 0 to approximately -8 eV/atom, and forces ranging from ~-75 to 75 eV/Å. The root mean square force and energy errors for the MLIP are 145 meV/atom and 332 meV/Å, respectively, resulting in a normalized error of 167. Although the energy error may be high compared to some NN-based MLIPs, the normalized error is comparable to those of single element metallic materials [24]. The discrepancy in energy and force errors in comparison to other MLIPs can be attributed to greater energy and force diversity within the training set. As mentioned previously, the data was primarily generated via a GA, which can yield a high proportion of non-energetically favorable structures. Other MLIPs which predominately employ sampling methods like AIMD may exhibit lower error levels because (i) the training and testing data tend to be highly correlated and/or (ii) AIMD simulations may be initialized with stable/meta-stable structures, which may constrain structural evolution and limit dataset diversity.

In addition, LAMMPS energy evaluations were conducted on approximately 200 structures with varied compositions using both the MLIP and Vashishta potential, as depicted in Figs. 3(c) and 3(d), respectively. The MLIP demonstrates high accuracy in energy predictions compared to DFT. Notably, the MLIP's performance remains consistent across compositions, whereas the Vashishta potential experiences large errors when evaluating structures with non-equimolar compositions. This performance disparity can be attributed to differences in model development practices. The MLIP was developed using a data driven approach aimed at learning the physics of the Si-C system, whereas the Vashishta potential was formulated for stoichiometric SiC only.

Furthermore, the MLIP's ability to accurately evaluate non-stoichiometric structures highlights its ability to capture some of the underlying physics. This capability is crucial for modeling material phenomena like melting [73], sublimation [63,74], etc., where non-equimolar compositions are common. Beyond modeling SiC in MD simulations, consistent predictive performance across the Si-C system bolsters the MLIP's use as a surrogate model in GAs, where it can be used in place of costly *ab initio* calculations to expedite the discovery of next-generation materials.

In further validation of the MLIP's capability to capture same-atom interactions, along with C-Si bonds, the pairwise energy interactions for Si-Si, C-Si, and C-C were evaluated and compared against Vashishta predictions (Fig. 4). Overall, the MLIP captures key attributes, including the presence of repulsive and attractive regions for each interaction and the correct ordering of energy wells. Specifically, the energy well depths, from deepest to shallowest follow the sequence C-C, C-Si, and Si-Si, in agreement with our DFT calculations. In contrast, Vashishta predicts completely repulsive same-atom interactions. This distinction helps explain the MLIPs greater ability to model non-equimolar structures as compared to the Vashishta potential. Furthermore, accurately representing these attributes is essential for effective modeling, as the presence of attractive regions plays a crucial role in describing complex behaviors, such as phase separation.

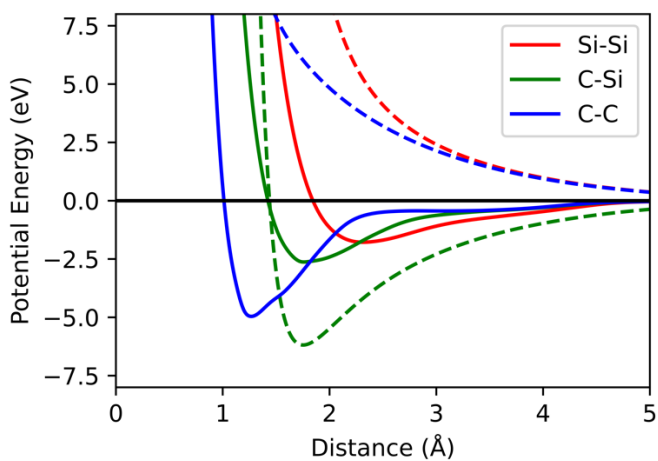

Figure 4: Comparison of MLIP (solid lines) and Vashishta (dotted lines) predicted pair interaction energy curves.

## 3.2 Property prediction

Along with capturing energy and force errors (the data that the model was trained on), a successful MLIP must also be able to accurately predict properties of interest. For a hard ceramic like SiC, with applications in structural and refractory fields, the MLIP's ability to accurately predict mechanical and thermal properties is paramount.

### 3.2.1. Fundamental mechanical properties

The first step towards validating the MLIP involves calculating fundamental material properties such as cohesive energy ($E_c$), lattice constants ($a, b, c$), elastic constants ($C_{ij}$), and bulk ($K$) and shear moduli ($G$). The MLIP's predictions are shown in Table II along with comparisons to values obtained from the Vashista potential, DFT, and experiments.

**Table II: Comparison of mechanical properties for common SiC polytypes from MLIP, DFT, experimental values, and the Vashishta potential**

| Polytype | Property | Unit | DFT | MLIP | Vashishta | Experiment |
|---|---|---|---|---|---|---|
| SiC-3C (zinc-blende) | $a = b = c$ | Å | 4.38 | 4.40 | 4.36 | 4.36 [75] |
| | $E_c$ | eV/atom | -7.53 | -7.40 | -6.34 | -6.34 [76,77] |
| | $C_{11}$ | GPa | 384 | 435 | 390 | 390 [78] |
| | $C_{12}$ | GPa | 127 | 118 | 142.6 | 142 [78] |
| | $C_{44}$ | GPa | 241 | 210 | 136.9 | 256 [78] |
| | $K$ | GPa | 213 | 224 | 225 | 225 [79] |
| | $G$ | GPa | 187 | 187 | 131.5 | 192 [79] |
| SiC-2H (wurtzite) | $a = b$ | Å | 3.09 | 3.10 | 3.06 | 3.08 [80] |
| | $c$ | Å | 5.07 | 5.08 | 5.02 | 5.05 [80] |
|         | $E_c$ | eV/atom | -7.53 | -7.33 | -6.32 | - |
|         | $C_{11}$ | GPa | 495 | 525 | 416 | - |
|         | $C_{33}$ | GPa | 534 | 553 | 379 | - |
|         | $C_{12}$ | GPa | 101.1 | 158 | 158 | - |
|         | $C_{13}$ | GPa | 49.2 | 135 | 149 | - |
|         | $C_{44}$ | GPa | 151.8 | 149 | 126.4 | - |
|         | $C_{66}$ | GPa | 196.8 | 183 | 128.7 | - |
|         | $K$ | GPa | 214 | 273 | 236 | - |
|         | $G$ | GPa | 185 | 173 | 125.8 | - |
|         | $a = b$ | Å | 3.08 | 3.10 | 3.08 | 3.08 [81] |
|         | $c$ | Å | 15.18 | 15.25 | 15.10 | 15.12 [81] |
|         | $E_c$ | eV/atom | -7.53 | -7.38 | -6.33 | - |
|         | $C_{11}$ | GPa | 485 | 498 | 406 | 501 [82] |
|         | $C_{33}$ | GPa | 535 | 513 | 415 | 553 [82] |
| SiC-6H  | $C_{12}$ | GPa | 104.3 | 123 | 142.8 | 111 [82] |
|         | $C_{13}$ | GPa | 51.1 | 106 | 140.6 | 52 [82] |
|         | $C_{44}$ | GPa | 160.8 | 166 | 131.1 | 163 [82] |
|         | $C_{66}$ | GPa | 190.1 | 188 | 132.2 | 195 [82] |
|         | $K$ | GPa | 213 | 242 | 230 | 220 [82] |
|         | $G$ | GPa | 187 | 181 | 132.4 | 191[a] |

[a]Calculated using experimental values from [82]

The MLIP's excellent ability to model the three polytypes is evident as the values are in close agreement to the DFT predicted values. Moreover, the MLIP captures all three polytype's shear moduli and the $C_{33}$ elastic constant for the hexagonal polytypes with a high degree of accuracy when compared to DFT and experimental data (if available), whereas Vashishta experiences larger errors.

It is worth noting that disparities between experimental and DFT cohesive energy values can stem from several factors, including size effects, experimental controls like sample purity and preparation, and quantum mechanical effects difficult to account for during simulation. Nonetheless, emphasis was placed on reproducing DFT predictions as they constitute the training data.

### 3.2.2 Equations of state

Expanding beyond property prediction, molecular dynamics simulations were performed to derive equations of state. First, the high-pressure behavior of SiC-3C was investigated using the MLIP, Vashishta, and DFT. In this analysis, the ground state structure underwent ±10% volumetric strain.

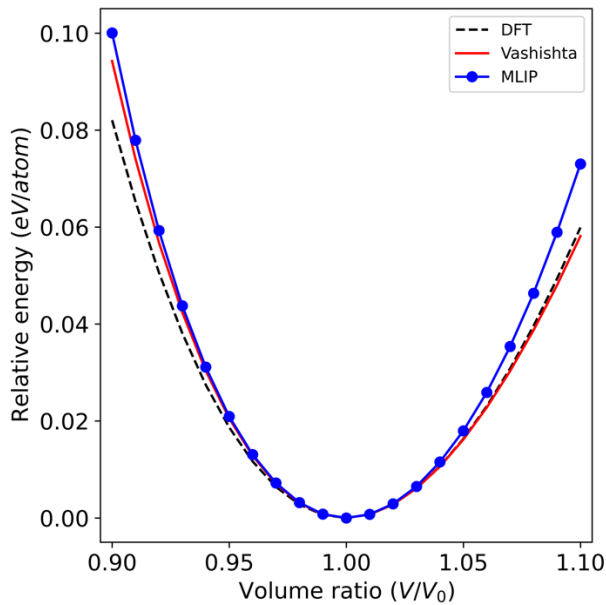

Figure 5: Relative energy per atom versus volume ratio for DFT, Vashishta, and MLIP.

To address disparities in cohesive energy predictions, the predicted energies were centered at the origin and then plotted as a function of volume-ratio (Fig. 5). Both the MLIP and Vashishta relative energy predictions are in overall agreement with DFT. However, the discrepancies in relative energies tend to increase for both the MLIP and Vashishta as the structure is strained. Achieving comparable relative energies to DFT is crucial, as these energy differences influence computed forces during classical atomistic simulations.

It is important to note that the MLIP was not fit to specific properties. Instead, the MLIP has successfully learned structure-energy relationships within the training data, without explicit knowledge of SiC deformation behavior, thereby demonstrating the transferability of the MLIP. In contrast, many potentials are calibrated to specific properties. For example, Vashishta potential's parameters were determined using physical properties, including the $C_{11}$ elastic constant and bulk modulus of SiC-3C [58]. Furthermore, while MLIPs are not fit to specific properties, many MLIPs are trained explicitly on strained configurations, similar to those assessed in a LAMMPS simulation for the determination of elastic constants. While this approach may yield near *ab initio* predictions as in the Vasishta potential (Fig. 5), it does not truly reflect an MLIP's generalizability.

From the same simulations used to generate Fig. 5, volume and pressure data were collected and used to fit a Murnaghan EOS, (described in Eqs. (2) and (3)) with the parameters presented in Table III. While the MLIP's fitted EOS overpredicts the bulk modulus ($K_0$) compared to Vashishta, the MLIP captures the pressure-volume relationships ($K_0'$) to a much better degree than Vashishta.

**Table III: Murnaghan Equation of State Parameters**

|  | DFT | MLIP | Vashishta |
|---|---|---|---|
| $K_0$ (GPa) | 211.3 | 261.0 | 225.0 |
| $K_0'$ | 3.76 | 3.97 | 6.23 |

To further validate the MLIP, we compared the MLIP's volume-pressure (Fig. 6(a)) and volume-temperature (Fig. 6(b)) relationships for SiC-3C at various temperatures ($T = 5, 300, \text{and } 1200 \text{ K}$) and pressures ($P = 0, 10, 20, \text{and } 70 \text{ GPa}$), respectively, to DFT calculations [83]. The MLIP aligns well with the DFT values, as the volume of the unit cell has an inverse relationship with pressure (Fig. 6(a)), and a positive relationship with temperature (Fig. 6(b)). The volume-pressure (Eq. (4)) and volume-temperature (Eq. (5)) equations of state were fit and compared to literature DFT values [83] in Table IV. The largest source of discrepancies in the previous volume relationships are the constant terms which stem from variations in lattice constant predictions shown earlier in Table 1.

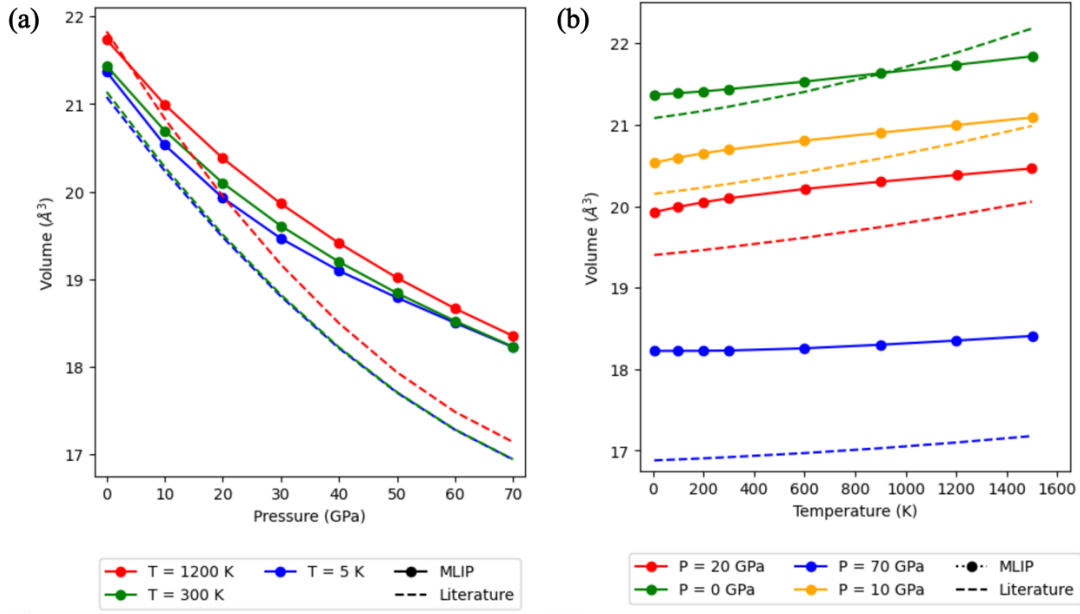

Figure 6: (a) Volume versus pressure at different temperatures and (b) volume versus temperature at different pressures.

Table IV: Comparison of MLIP and literature volume-pressure and volume-temperature EOS parameters.

|  | T (K) | MLIP | | | Literature [83] | | |
|---|---|---|---|---|---|---|---|
|  |  | $V_0$ | $V_0'$ | $V_0''$ | $V_0$ | $V_0'$ | $V_0''$ |
| $V(P)$ | 5* | 21.29 | $-7.29 \times 10^{-2}$ | $4.30 \times 10^{-4}$ | 21.08 | $-8.84 \times 10^{-2}$ | $4.18 \times 10^{-4}$ |
|  | 300 | 21.39 | $-6.97 \times 10^{-2}$ | $3.58 \times 10^{-4}$ | 21.14 | $-9.02 \times 10^{-2}$ | $4.32 \times 10^{-4}$ |
|  | 1200 | 21.70 | $-7.12 \times 10^{-2}$ | $3.39 \times 10^{-4}$ | 21.83 | $-1.05 \times 10^{-1}$ | $5.43 \times 10^{-4}$ |
|  | P (GPa) | MLIP | | | Literature [83] | | |
|  |  | $V_0$ | $V_0'$ | $V_0''$ | $V_0$ | $V_0'$ | $V_0''$ |
| $V(T)$ | 0 | 21.36 | $2.58 \times 10^{-4}$ | $0.417 \times 10^{-7}$ | 21.08 | $4.09 \times 10^{-4}$ | $2.17 \times 10^{-7}$ |
|  | 10 | 20.55 | $4.76 \times 10^{-4}$ | $-0.80 \times 10^{-7}$ | 20.15 | $3.77 \times 10^{-4}$ | $1.20 \times 10^{-7}$ |
|  | 20 | 19.94 | $5.515 \times 10^{-4}$ | $-1.15 \times 10^{-7}$ | 19.40 | $3.00 \times 10^{-4}$ | $0.92 \times 10^{-7}$ |

| 70 | 18.22 | $0.22 \times 10^{-4}$ | $0.70 \times 10^{-7}$ | 16.88 | $1.17 \times 10^{-4}$ | $0.55 \times 10^{-7}$ |

*MD simulations were conducted at 5 K, whereas literature values are for 0 K.

### 3.2.3 High temperature behavior

In the previous section, the MLIP predictive performance was demonstrated through practices commonly employed by researchers while validating their MLIPs and empirical potentials [58,84–86]. In this section, the MLIPs performance on phenomena proven difficult to model with empirical potentials, such as thermal decomposition and surface reconstruction, are summarized. The aim of these studies is to showcase the MLIP's ability to model complex phenomena and spur future investigations into SiC behaviors (e.g., silicon carbide epitaxy). It must be emphasized that the phenomena modeled may not be directly represented in the training data, underscoring the MLIPs ability to capture underlying physics of the Si-C system.

**Thermal decomposition**

To assess the thermal decomposition behavior of SiC as predicted by the MLIP, melting simulations were performed using the coexistence method. The thermal decomposition temperature was determined by monitoring the prevalence of cubic diamond structures during the simulation. The MLIP predicts a decomposition temperature of 1825 K. At temperatures exceeding this threshold, crystalline SiC undergoes a peritectic reaction, where silicon sublimes leaving behind solid carbon. This peritectic reaction is consistent with the SiC phase diagram [60] and experimental studies [73,87–89]. Moreover, the predicted temperature is in excellent agreement with experimental observations, where thermal decomposition has been seen to occur from 1200 K – 2100 K [60–63,87,88,90,91].

To further examine this peritectic reaction, a carbon-terminated SiC-3C (001) surface was equilibrated at 2000 K. The simulation involved a 12x12x26 supercell, totaling 29952 atoms, with periodic boundary conditions in the x and y directions, and free surface boundaries in the z direction. As seen in Figure 7(a), the MLIP effectively captures the separation of silicon (brown atoms) and carbon (grey atoms) resulting from silicon sublimation.

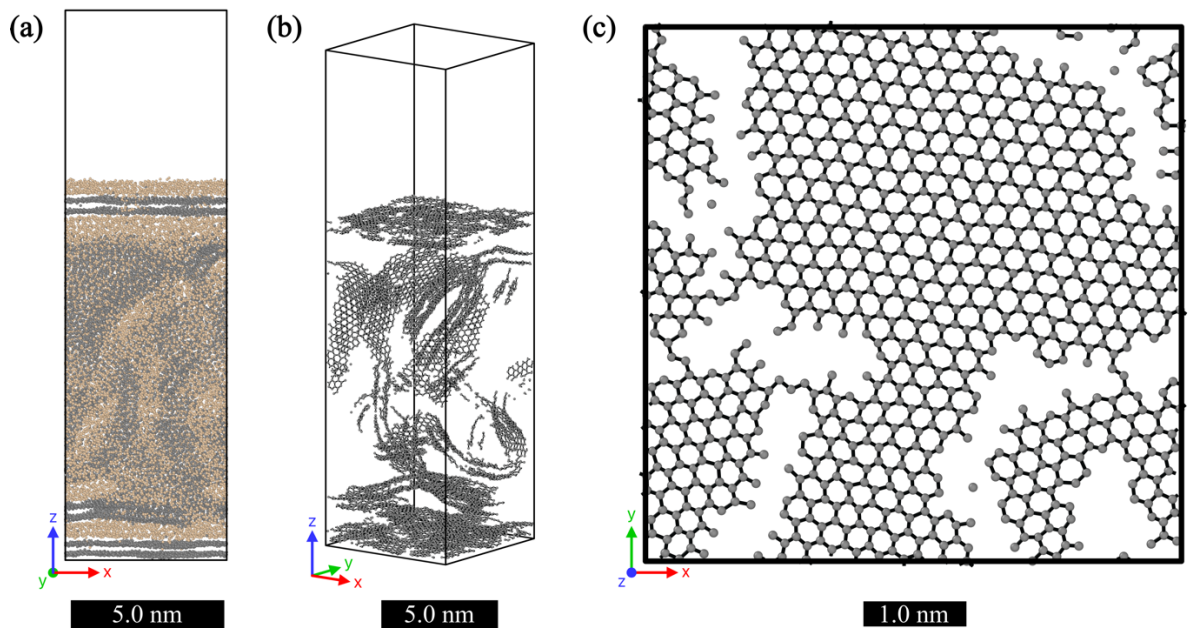

Figure 7: (a) Separation of silicon (brown atoms) and carbon (grey atoms), (b) demonstrating the formation of graphitic structures, (c) zoomed in view of graphene sheet formed due to silicon sublimation (corresponds to bottom sheet of carbon (grey atoms) in figure 7a).

Additionally, polyhedral template matching [92] in Ovito [93] facilitated the identification of separated carbon regions as graphitic structures (Fig. 7b). Specifically, graphene sheets formed at the upper and lower surfaces with a characteristic honeycomb pattern (Fig. 7c). These findings are supported by studies on silicon carbide decomposition and graphene growth [62], where silicon carbide decomposition has been explored as a method for graphene synthesis [61,63,94]. The structure of these graphene sheets was quantitatively validated via bond length analysis of the carbon rings within the honeycomb structures. The average carbon ring bond length as predicted by the MLIP is 142.5 pm, closely aligning with DFT and experimental studies (~142 pm [95] and ~141.8 pm [62], respectively).

The MLIPs capability to model thermal decomposition is seen as an advancement, where the Vashishta potential predicts a crystalline to liquid transition with no graphene formation [96,97], and another popular potential, Tersoff [98], requires the removal of a silicon atom to simulate thermal decomposition [99]. The MLIPs capability to capture graphene growth from silicon carbide decomposition presents opportunities for a deeper atomic level understanding of graphene synthesis. Moreover, it can enable insights on longer time and length scales than those achievable through *ab initio* methods and facilitates the assessment of atomic level behaviors that are challenging to observe experimentally.

**Surface reconstruction**

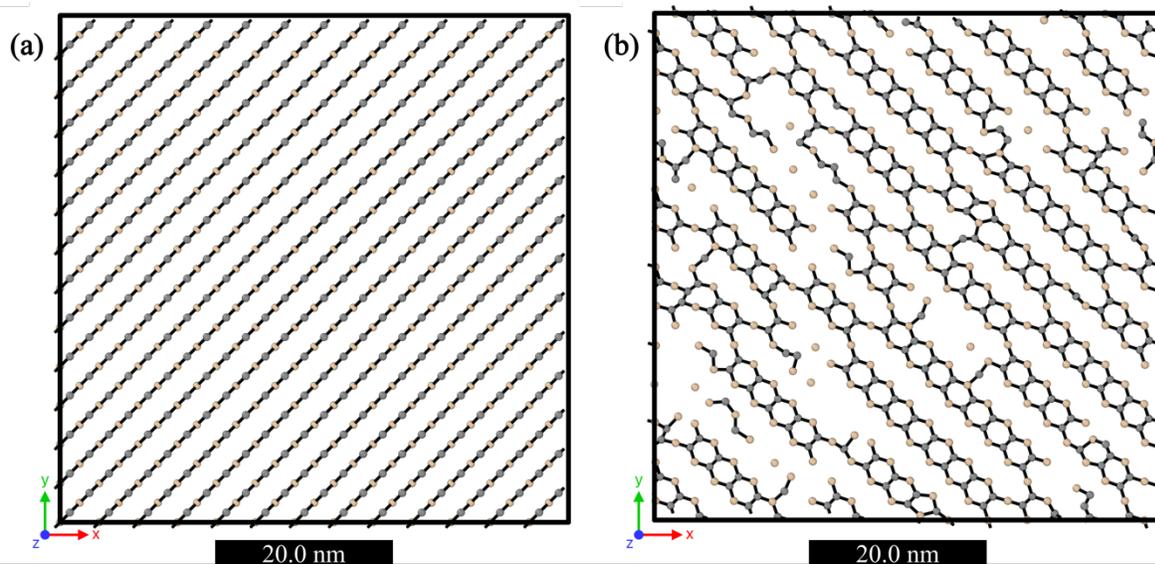

Figure 8: (a) Initial and (b) final images of a carbon-terminated SiC-3C (001) surface equilibrated at 1200 K.

The MLIP's performance to predict surface reconstruction was evaluated by building a model with a carbon terminated SiC-3C (001) surface and equilibrating to 1200 K. The simulation used a 13x13x6 supercell, totaling 8112 atoms, with periodic boundaries in the x and y directions, and free surface boundaries in the z direction. Shown in Figs. 8(a) and 8(b) are initial and final images of the SiC surface, respectively.

As depicted above, the surface atoms have undergone surface reconstruction. More specifically, the carbon atoms, initially bonded to the silicon atoms in a layer below, have formed C-C dimers. This reconstruction aligns with existing literature, notably demonstrating the dimer-row reconfiguration [100–102], commonly observed in the simulated carbon-terminated SiC-3C configuration. Moreover, measurements of C-C and C-Si bond lengths are ~1.47 Å and ~1.84 Å, respectively. These values closely resemble literature findings of 1.36 Å for C-C and 1.83-1.91Å for C-Si [100,103,104].

The above two examples further demonstrate the range of applicability of the MLIP, where complex, difficult to model phenomena, were effectively modeled, in agreement with *ab initio* and experimental findings. Moreover, the MLIPs excellent performance in modeling these behaviors invites further investigation into material interfaces within the Si-C system like SiC/graphene.

## Conclusion

A MLIP has been developed for molecular dynamics simulations of the Si-C system, achieving excellent accuracy relative to *ab initio* in reproducing mechanical properties and capturing the physics of the Si-C system under different thermodynamic conditions. The MLIP was trained using the linear regression based $UF^3$ development package. $UF^3$ was chosen for its computational efficiency and robustness in capturing the physics of a material system with minimal data compared to other approaches such as NNs. To enhance MLIP performance, modifications were made to the $UF^3$ featurization process, including increased model fitting parameters around pair interaction energy wells. This modification resulted in significant improvements in MLIP performance, particularly in the predictive accuracy of mechanical properties.

The developed MLIP effectively learned the underlying structure-energy relationships within the Si-C system, experiencing normalized errors comparable to single element metal MLIPs, and demonstrated excellent predictive performance across stoichiometries. This robust performance was further confirmed through extensive evaluation of fundamental properties (e.g., elastic constants, cohesive energy) and volume-pressure/temperature relationships, all of which closely aligned with *ab initio* predictions.

Furthermore, the MLIP's predictive capabilities as well as range of applicability were highlighted through the simulation of complex SiC behaviors, including thermal decomposition and surface reconstruction. The MLIP accurately predicted SiC's peritectic reaction at a temperature (1825 K) commensurate with experiments (1200-2100 K), and carbon-dimer surface reconstruction. These results highlight the MLIPs suitability for modeling complex phenomenon, pertinent to practical applications of SiC. Ultimately, the approach adopted here with genetic algorithms and UF3 potentials is shown to be a promising approach for developing MLIPs for ceramics with complex structures and behaviors.

## Declarations

**Funding information**

The authors acknowledge University of Florida Research Computing for providing computational resources and support that have contributed to the research results reported in this publication. URL: http://www.rc.ufl.edu.

**Conflicts of interest**

The authors have no relevant financial or non-financial interests to disclose.

**Data availability**

The data used to train the model; python scripts used during the featurization and training; and the final model can be found at https://github.com/SubhashUFlorida/SiC-MLIP.

**Author contributions**

Michael MacIsaac: Conceptualization, Methodology, Software, Formal Analysis, Visualization, Writing – Original Draft.

Salil Bavdekar: Methodology, Formal Analysis, Data Curation, Writing – Review & Editing.

Douglas Spearot: Conceptualization, Software, Resources, Supervision, Writing – Review & Editing.

Ghatu Subhash: Conceptualization, Resources, Writing – Review & Editing, Supervision, Project Administration.